\documentstyle[12pt]{article}

\def\thefootnote{\fnsymbol{footnote}}

\newcommand{\eq}{\begin{equation}}
\newcommand{\en}{\end{equation}}
\newcommand{\eqa}{\begin{eqnarray}}
\newcommand{\ena}{\end{eqnarray}}

\newcommand{\NP}[1]{Nucl.\ Phys.\ {\bf #1}}
\newcommand{\PL}[1]{Phys.\ Lett.\ {\bf #1}}

\newcommand{\PR}[1]{Phys.\ Rev.\ {\bf #1}}
\newcommand{\PRL}[1]{Phys.\ Rev.\ Lett.\ {\bf #1}}

\begin{document}
\begin{titlepage}
\vskip0.5cm
\begin{flushright}
DFTT 2/97\\
HUB-EP-97/1\\
\end{flushright}
\vskip0.5cm
\begin{center}
{\Large\bf Universal Amplitude Ratios}
\vskip 0.3cm
{\Large\bf in the 3D Ising Model}
\end{center}
\vskip 1.3cm
\centerline{
M. Caselle$^a$\footnote{e--mail: caselle~@to.infn.it}
 and M. Hasenbusch$^b$\footnote{e--mail: hasenbus@birke.physik.hu-berlin.de}}
 \vskip 1.0cm
 \centerline{\sl  $^a$ Dipartimento di Fisica
 Teorica dell'Universit\`a di Torino}
 \centerline{\sl Istituto Nazionale di Fisica Nucleare, Sezione di Torino}
 \centerline{\sl via P.Giuria 1, I-10125 Torino, Italy}
 \vskip .4 cm
 \centerline{\sl $^b$ Humboldt Universit\"at zu Berlin, Institut f\"ur Physik}
 \centerline{\sl Invalidenstr. 110, D-10099 Berlin, Germany}
 \vskip 1.cm

\begin{abstract}
We present a high precision Monte Carlo study of various universal amplitude
ratios of  the three dimensional Ising spin model. Using state of the art simulation
techniques we studied the model close to criticality in both phases.
Great care was taken to control systematic errors due to finite size effects
and correction to scaling terms. We obtain  $C_+/C_-=4.75(3)$, 
$f_{+,2nd}/f_{-,2nd}=1.95(2)$ and $u^*=14.3(1)$. 
Our results are compatible with those obtained by field theoretic methods  
applied to the $\phi^4$ theory  and high and low temperature 
series expansions of the Ising model.
The mismatch with a previous Montecarlo study by Ruge et al. remains to be 
understood. 

\vskip0.2cm
\end{abstract}
\end{titlepage}

\setcounter{footnote}{0}
\def\thefootnote{\arabic{footnote}}

\section{Introduction}
 In the neighbourhood of a second order phase transition
 various quantities display
 a singular behaviour. In this limit most of the microscopic features which
 characterize a given model become irrelevant and models which 
 differ at the microscopic level may share the same singular behaviour. This is
 the basis of the concept of universality.
 The first, well known, consequence of universality is that
different models belonging to the same universality class share the same
 critical indices. However the hypothesis of 
 universality of the various scaling functions has much stronger implications
  and it is possible to show that models belonging to the same universality
  class are also
 characterized  by the same values of some critical-point amplitude 
combinations~\cite{ahp}. Let us see a simple example.
Near the critical point the correlation length $\xi$ diverges as
 \begin{eqnarray}
 \xi &\sim& f_+ \;\; t^{-\nu}  ;  \;\;\;\;\;\;\; t > 0 \nonumber \\
 \xi &\sim& f_- \; (-t)^{-\nu} ;  \;\; t < 0
 \end{eqnarray}
 with 
 \begin{equation}
  t = \frac{T-T_c}{T_c}
 \end{equation}
 where $T$ is the temperature and $T_c$  the critical temperature. 
Different models in the same universality class share not only
 the same critical exponent $\nu$, but also the same 
 dimensionless combination of critical amplitudes  $f_+/f_-$.
This is particularly relevant from the experimental point of view, since in
general critical amplitudes are more easily detectable than  critical indices
and allow a simpler identification of the universality class. In fact the
variations of the critical indices between different universality classes are
in general rather small, while the amplitude ratios may vary by large amounts.
In this paper we shall be in particular interested in the universality class of
the three dimensional Ising model which has several interesting experimental
realizations, ranging from the binary mixtures to the liquid vapor transitions.

The  two standard approaches to the evaluation of these
 amplitudes ratios in the Ising case are the use of field theoretic methods 
applied to the 
$\phi^4$ theory \cite{blz}-\cite{munster2}, and the
extrapolation to criticality 
of low and high temperature
series expansions on various lattices \cite{lf}. All these estimates are in
general in rather good agreement among them (for a comparison and a discussion
see  sect.5). 

In order to obtain results from Montecarlo
simulations  relevant for the scaling limit
we have to control both finite size effects as well as corrections
to scaling. 
This means that the linear lattice sizes $L$ have to be chosen such that 
$L >> \xi$ while  $\xi \rightarrow \infty$ as $\beta \rightarrow \beta_c$. 
In practice one has to carefully check which factor of 
$L /\xi$ is required to obtain results sufficiently close to the thermodynamic
limit. While in the high temperature phase $L /\xi \approx 7$ turns out to 
be sufficient to 
give thermodynamic limit results within numerical accuracy,  in the low temperature 
phase this factor has to be doubled at least.  
The value of $\xi$ that can be reached, and hence the control of corrections to 
scaling, is limited by the CPU time available for the study.  In the present paper
the largest correlation length is $\xi=11.884(9)$
in the high temperature phase
and $\xi=6.208(18)$ in the low temperature phase. 

One also should note that simulations in the low temperature phase are
considerably more difficult than those in the high temperature phase of the model. 
In the low temperature phase conceptual as well as practical problems
caused by spontaneous symmetry breaking arise. Furthermore  the determination 
of the correlation length is complicated by the occurrence of secondary correlation 
lengths which are close to the leading one.

For all these reasons 
only in these last years there have been some attempts to measure these 
ratios in Montecarlo simulations~\cite{kielxi}. However the results are rather
puzzling. For instance, in the case of the ratio $f_+/f_-$ discussed above,
the Montecarlo estimate, which
is $f_+/f_-=2.06(1)$ \cite{kielxi}, disagrees with the one obtained with
strong/weak coupling series $f_+/f_-=1.96(1)$~\cite{lf}, while the
field theoretical estimate $f_+/f_-=2.013(28)$~\cite{munster2} lies in
between the two. 

The aim of our work is to show that Montecarlo estimates of the amplitude
ratios can indeed be competitive with other approaches. To this end
 we have elaborated
 a technique to directly extract the various amplitude ratios, without 
evaluating  the single amplitudes thus avoiding all the uncertainties
related to the critical indices. We shall discuss this point in sect.5 below.
 Besides this, we
have devoted a great care throughout the paper to  keep under control
systematic errors due to 
finite size effects and corrections to scaling.
Finally we have used state of art simulation techniques to obtain 
 high precision estimates of the observables
 near the critical point, in both phases.
The simulations in the high temperature phase have been performed using Wolff's
single cluster algorithm.
Here the improved estimators give a great boost to the 
accuracy of the results. 
However in the low temperature phase, due to the finite magnetization,
the improved
estimators of the cluster-algorithm are of little help. Hence we simulated here 
with a multispin coding implemented Metropolis-like algorithm. 

As we shall see our results are comparable in precision and agree with
 the most recent field-theoretic and strong/weak coupling estimates, while 
they are incompatible with the MC results of Ruge et al.
~\cite{kielxi}. 
A more detailed comparison of the data might be helpful to understand this 
discrepancy. 

This paper is organized as follows:
In sect.2 we have collected some informations on the three dimensional Ising
model and on the observables that we study. In sect.3 we discuss the details of
the simulation, while in sect.4 we analyse the scaling behaviour of the 
measured  quantities: magnetization, susceptibility and correlation lengths.
In sect.5 we study the amplitude ratios and compare our results with other
existing estimates and with the experiments. Finally sect.6 is devoted to some
concluding remark.

\section{General Setting}
\subsection{The Model}

We study the Ising spin model in three dimensions on a simple cubic lattice. 
The action is given by
\eq
 S_{spin} = - \beta \sum_{<n,m>} s_n s_m \; , 
\label{Sspin}
\en
where
 the field variable $s_n$ takes the values $-1$ and $+1$;
 $~~n\equiv(n_0,n_1,n_2)$ labels the sites of the lattice and the notation 
$<n,m>$  indicates that the sum is taken on  nearest neighbour sites 
only. The coupling $\beta$ is defined as  $\beta\equiv \frac{1}{kT}$, hence
the reduced temperature $t$ can be written as
\eq
t=\frac{\beta_c-\beta}{\beta}~~~,
\label{tbeta}
\en
where $\beta_c\equiv \frac{1}{kT_c}$.
We shall consider in the following $n_1$ and $n_2$ as ``space'' directions
and   $n_0$  as the ``time'' direction and shall sometimes denote the time
coordinate $n_0$ with $\tau$.
We  always consider lattices of 
equal extension $L$ and periodic boundary conditions in all three directions.

\subsection{The observables}

\subsubsection{Magnetisation}
The magnetization of a given configuration is defined as: 
\begin{equation}
m = \frac{1}{V} \sum_i s_i \;\;,
\end{equation}
where $V\equiv L^3$ is the volume of the lattice. 
However, in a finite volume the $Z_2$ symmetry of the model can not be broken
  for any nonzero temperature. Hence the expectation value 
of $m$ vanishes. 

In order to obtain the magnetization of the model in the low temperature phase
one should add a magnetic field $h$ in order to break 
the symmetry. Then one should first take
the thermodynamic limit at finite magnetic field and then take the limit
of vanishing magnetic field.
However it is difficult to follow this route in a numerical study. 

As an alternative Binder and Rauch \cite{binder} suggested to 
simulate the finite lattices at vanishing  external field 
and study the quantity
\begin{equation}
<m>~\equiv~\lim_{L\to\infty} \sqrt{<m^2>}
\end{equation}

However it turns out that this is not the best choice. In fact
 this observable is affected by strong finite size
effects~\cite{mtis} which would require very large lattices to obtain reliable
estimates of the infinite volume magnetization. 
It has been recently observed~\cite{tb} that a much more stable observable is:

\begin{equation}
<m>~\equiv ~\lim_{L\to\infty} <|m|>   \;\;\; .
\end{equation}

The finite size behaviour of this observable, was carefully 
studied in~\cite{tb} where it was shown that the asymptotic, infinite volume,
value is reached for lattices of size $L>\sim 8\xi$, where $\xi$ denotes the
correlation length. In our simulations we always used lattice
sizes much larger than this threshold.

Close to the critical temperature, the magnetization is supposed to scale as

\begin{equation}
\label{defb}
 <m> \sim \; B \; (-t)^{\beta} ;  \;\; t < 0
\end{equation}

where the critical exponent $\beta$ should
not be confused with the inverse temperature. 

\subsubsection{Magnetic susceptibility}
The susceptibility gives the response of the magnetization 
to an external magnetic
field. 
\begin{equation}
\chi = \frac{\partial <m> }{\partial H}
\end{equation}
One easily derives that the magnetic susceptibility can be expressed in 
terms of moments of the magnetization as follows: 
\begin{equation}
\chi = V \left( <m^2> - <m>^2 \right)~~~~~~.
\end{equation}
Close to the critical temperature the magnetic susceptibility is supposed
to scale as
\begin{eqnarray}
\label{defc}
\chi &\sim& C_+ \;\; t^{-\gamma}  ;  \;\;\;\;\;\;\; t > 0 \nonumber \\
\chi &\sim& C_- \; (-t)^{-\gamma} ;  \;\; t < 0~~~~~~~.
\end{eqnarray}
\subsubsection{Exponential correlation length.}
We consider the decay of so called time-slice
correlation functions. The magnetization of a time slice is given by
\begin{equation}
 S_{n_0} = \frac{1}{L^2}\sum_{n_1,n_2} s_{(n_0,n_1,n_2)} \;\;\; .
\label{timeslice}
\end{equation}
Let us define the correlation function 
\begin{equation}
G(\tau) = \sum_{n_0} \left\{ \langle
 S_{n_0} S_{{n_0}+\tau} \rangle - \langle S_{n_0} \rangle^2 \right\}\;\;\; .
\end{equation}
The large distance behaviour of $G(\tau)$ is given by
\begin{equation}
G(\tau) \propto \exp(-\tau/\xi) \;\;\; ,  
\end{equation}
where $\xi$ is the exponential correlation length. 

Close to criticality the behaviour of the correlation length is governed 
by the scaling laws
\begin{eqnarray}
\label{deff}
\xi &\sim& f_+ \;\; t^{-\nu}  ;  \;\;\;\;\;\;\; t > 0 \nonumber \\
\xi &\sim& f_- \; (-t)^{-\nu} ;  \;\; t < 0~~~~~~~.
\end{eqnarray}

\subsubsection{Second moment correlation length.}
The square of the second moment correlation length is defined for a generic
value of the spacetime dimensions $d$ by
\begin{equation}
 \xi_{2nd}^2 = \frac{\mu_2}{2 d \mu_0} \;\;,
\label{mu2}
\end{equation}
where 
\begin{equation}
\mu_0 = \lim_{L \rightarrow \infty}\; \frac{1}{V} \; \sum_{m,n} \;
\langle s_m s_n \rangle_c
\end{equation}
and 
\begin{equation}
\mu_2 = \lim_{L \rightarrow \infty}\; \frac{1}{V} \;  \sum_{m,n} \;
(m - n)^2 \langle s_m s_n \rangle_c \;.
\end{equation}
The connected part of the correlation function is given by
\begin{equation}
 \langle s_m s_n \rangle_c = 
\langle s_m s_n \rangle -  \langle s_m  \rangle ^2
\end{equation} 

This estimator for the correlation length  is very popular
since its numerical evaluation (say in Montecarlo simulations) is simpler than
that of the exponential correlation length. Moreover it is the length scale
which is directly observed
in scattering experiments. However it is important
to stress that it is not  exactly equivalent to the exponential correlation
length. 
The relation between the two can be obtained as follows.
Let us write
\begin{eqnarray}
\mu_2 &=& \frac{1}{V}\; \sum_{m;n} \; (n - m)^2 \;\; \langle s_m s_n \rangle_c
\nonumber \\
   &=& \frac{1}{V} \; \sum_{n;m} \; \sum_{\mu=0}^{d-1} \;
  (n_{\mu} - m_{\mu})^2 \;\;  \langle s_m s_n \rangle_c
  \nonumber \\
   &=& \frac{d}{V} \; \sum_{n;m} \; (n_0 - m_0)^2  \;\;
\langle s_m s_n \rangle_c~~~.
\end{eqnarray}
Due to the exponential decay of the correlation function this sum is certainly
convergent and we 
can commute the spatial summation with the summation over
configurations so as to obtain
\begin{equation}
\mu_2 = d \sum_{\tau=-\infty}^{\infty} \; \tau^2 \;\; \langle S_0 \; S_\tau 
 \rangle_c
\end{equation}
with $S_{n_0}$ given by eq.(\ref{timeslice}).
Analogously one obtains
\begin{equation}
\mu_0 = \sum_{\tau=-\infty}^{\infty} \; \langle S_0 \; S_\tau  \rangle_c
\end{equation}

If we now insert these results in eq.(\ref{mu2}), assume a multiple 
exponential decay 
\begin{equation}
\langle S_0 \; S_\tau  
\rangle_c \propto \sum_i \; c_i \; \exp(-|\tau|/\xi_i) \;\; , 
\end{equation}
and replace the summation by an integration over $\tau$  we get
\begin{equation}
\label{e24}
 \xi_{2nd}^2 = \frac12 \;
          \frac{\int_{\tau=0}^{\infty} {\mbox d}\tau \;\tau^2 
\; \exp(-\tau/\xi)}
                     {\int_{\tau=0}^{\infty} {\mbox d}\tau \; \exp(-\tau/\xi)}
           =   \frac{ \sum_i c_i \xi_i^3}{\sum_i c_i \xi_i} \;,
\end{equation}
which is equal to $\xi^2$ if only one state contributes. 
An interesting consequence of this analysis is that the difference from one of
the ratio $\xi/\xi_{2nd}$ gives an idea of the density of the lowest states of
the spectrum. If these are well separated the ratio will be almost one, while a
ratio significantly higher than one will indicate a denser distribution of
states.

The critical behaviour of $\xi_{2nd}$ is governed by the same critical index
$\nu$, so near the critical point we expect:
\begin{eqnarray}
\label{deff2}
\xi_{2nd} &\sim& f_{+,2nd} \;\; t^{-\nu}  ;  \;\;\;\;\;\;\; t > 0 \nonumber \\
\xi_{2nd} &\sim& f_{-,2nd} \; (-t)^{-\nu} ;  \;\; t < 0
\end{eqnarray}

\subsection{Critical indices}
Our aim is to obtain high precision estimates for some amplitude ratios. To
this end we need to use as input informations  the critical temperature 
$\beta_c$ and the values of the critical indices~\footnote{As a matter of
fact, for the actual determination of the amplitude ratios we only need to know
$\beta_c$ and $\theta$. The values of the other critical indices will only be
used in comparing our results with those obtained with the
series expansions.} defined above. 
We list in tab.1 some estimates for these 
quantities, obtained with field theoretical methods, strong coupling series and
Montecarlo simulations.

\begin{table}[h]
\label{literature1}
\caption{\sl Results for $\beta_c$ and for the critical indices 
 given in the literature}
\vskip 0.2cm
\begin{tabular}{|c|c|c|c|c|c|c|}
\hline
ref. &method & $\beta_c$ & $\gamma$ & $\nu$ & $\beta$ & $\theta$ \\
\hline
\cite{baker}& FT & &1.241(4) & 0.630(2)&0.324(6) & 0.496(3) \\
\cite{zl}& $\epsilon$-expansion &   
& 1.2390(25) & 0.6310(15) & 0.3270(15) &  0.51(3) \\
\cite{Parisi}& $d=3$ &  & 1.2405(15) & 0.6300(15) & 0.3250(15) & 0.50(2)\\
\cite{gupta1}& MCRG & 0.221652(3) && 0.624(1) & & 0.50-0.53\\
\cite{gupta2}& MCRG & 0.221655(1)(1) & & 0.625(1) & &  0.44 \\
\cite{blh}& MC,FS & 0.2216546(10)& 1.237(2)& 0.6301(8) & 0.3267(10)& 0.52(4)\\
\cite{lan}& MC & 0.2216576(22)& 1.239(7)& 0.6289(8) & 0.3258(44)& \\
\cite{tb}& MC & 0.2216544(3)& & & 0.3269(6) & 0.508(25) \\
\cite{nr}& HT &    & 1.237(2)& 0.6300(15) & & 0.52(3)\\
\cite{ge}& HT &    & 1.239(3)& 0.632(3) & & 0.52(3)\\
\cite{zj}& HT &    & 1.2385(25)& 0.6305(15) & & 0.57(7)\\
\cite{cfn}& HT &    & 1.2395(4)& 0.632(1) & & 0.54(5)\\
\hline
\end{tabular}
\end{table}

In general these values are  in rather good agreement among them,
despite the fact that they were obtained with very different methods.
As input parameters for our analysis we have decided to choose
the following values:
\eq
\beta_c=0.2216544(3)~,~~~\gamma=1.2390(15)~,~~~
 \nu=0.6310(15)~,\nonumber
\en
\eq
 \beta=0.3270(6)~,~~~ \theta=0.51(3).
\en
which are obtained  combining together the results of~\cite{tb}
and~\cite{zl} and satisfy the scaling relations. 
Let us stress however that our results are only slightly
affected by this choice~\footnote{The  systematic errors  
due to the uncertainties in the choice of $\beta_c$ and $\theta$ turn out to
be much smaller than the statistical fluctuations of our estimates.
In any case, both are taken  into account  in the  errors that we  quote 
in our final results.}.

\subsection{Amplitude ratios}
In  the following we shall be interested in these scaling functions:

\eq
\Gamma_{\chi}(t)\equiv \frac{\chi(t)}{\chi(-t)}~,~~~~~
\Gamma_{\xi}(t)\equiv \frac{\xi_{2nd}(t)}{\xi_{2nd}(-t)}~~~~(t>0),
\nonumber
\en

\begin{equation}
u(t)\equiv \frac{3~~\chi(t)}{\xi_{2nd}^3(t) m^2(t)}~~~~~~(t<0),
\label{u}
\end{equation}
\begin{equation}
\Gamma_c(t)\equiv \frac{\chi(t)}{\xi_{2nd}^3(t) m^2(-t)}~~~~~~(t>0)
\label{gammac}
\end{equation}
(notice
 the factor of three difference between the definitions of $\Gamma_c$ and
$u$).
While $\Gamma_\chi,\Gamma_\xi$ and $\Gamma_c$ mix low and high temperature
observables, $u$ only contains quantities evaluated
 in the broken symmetry phase. $\Gamma_c$ and $u$
are scale invariant thanks to the following 
scaling (and hyperscaling) relations among the
critical exponents:
\eq
\alpha+2\beta+\gamma=2~,~~~~~~~~~~~
d\nu=2-\alpha~~~.
\en
In particular, $u$
 plays the important role of a low temperature renormalized coupling
constant in  the study of the   $\phi^4$  theory directly
in $d=3$. 
 
It is important to notice that $\Gamma_c$ is related to the
ratio of two amplitude combinations (in which 
$A(t)$ denotes the specific heat):
\eq
 R_c\equiv
 \frac{\chi(t)~A(t)}{m^2(-t)}
 ~~~~ R_\xi\equiv{\xi_{2nd}(t)~A(t)^{1/3}} ~~~~~~(t>0)
\en
which have been widely
studied in the literature since they can be rather easily evaluated in
experiments. The relation is: $\Gamma_c=R_c/R_\xi^3$.
In the scaling limit these functions are related to the amplitudes
defined in eq.(\ref{defb},\ref{defc},\ref{deff},\ref{deff2}) as follows:
\eq
\lim_{t\to 0}~\Gamma_{\chi}(t) = \frac{C_+}{C_-}~~,~~~~~~~
\lim_{t\to 0}~\Gamma_{\xi}(t) =\frac{f_{+,2nd}}{f_{-,2nd}}~~,
\en
\eq
\lim_{t\to 0}~u(t)\equiv u^* = \frac{3~C_-}{f_{-,2nd}^3~ B^2}~~,~~~~~~~
\lim_{t\to 0}~\Gamma_c(t) =\frac{C_+}{f_{+,2nd}^3~ B^2}~~.
\en

 Finally we shall also be interested in evaluating the ratio:
\eq
\frac{\xi}{\xi_{2nd}}
\nonumber
\en
both above and below the
critical point.

\subsection{Series expansions}
A very powerful approach to the study of
the three dimensional Ising model is represented by the series
 expansions which lead to estimates for
several quantities near the critical point which are competitive with the most
precise Montecarlo simulations. In the following we shall compare our results
in the low temperature phase (which is the one in which simulations are more
difficult and  results are in general affected by stronger finite size
effects) 
with those obtained with series expansions with the twofold aim of testing the
reliability of our simulations and of  comparing the precision of the two
methods. The low temperature regime is also particularly interesting because
 recently these series have been extended up to very high 
orders~\cite{at1,at2,ge2}. Some
 informations on these series can be found in tab.13. 
In order to  extract from the series the 
estimates for the observables in
which we are interested and to quantify the uncertainty of such estimates 
 we  use the so called ``double biased
inhomogeneous differential approximants'' (IDA). The technique of IDA
 is  described in~\cite{lf,ida}, to which
we refer for notations and further details.  Following ref.~\cite{lf} we use 
the notation [K/L;M] for the approximants.  
In order to keep the fluctuations 
of the results under control, we have chosen to use double biased 
IDA~\cite{lf}, namely we  fix both the 
critical coupling $\beta_c$ and the critical index describing the critical
behaviour of the observable. As K and L vary we obtain several different IDA's
and correspondingly several different estimates of the observable.
As  we approach the critical point these estimates start to spread out, 
indicating that we are pushing the series toward its convergence threshold. 
The last problem is then to extract from this set of values the best estimate
and its uncertainty. Our choice in this respect is to neglect those IDA's
which fluctuate too wildly and treat the remaining
approximants on the same ground. To this end we determine
 the smallest interval that contains half of the results.
 The values that we shall
quote in the  tables below as our best estimates
 correspond to the centre of this interval,
 the first number in brackets
gives the half of the size of the interval. The second number in brackets gives
the error induced by the error of $\beta_c = 0.2216544(3)$  and
of the critical index used for biasing. Together they give an idea of the
uncertainty of the estimate. As we shall see, in the range of coupling in which
we are interested,  the uncertainty will be always
dominated by the spread of IDA's.

\section{The simulations}
\subsection{Simulations in the low temperature phase.}

We  simulated
the Ising spin model in its low temperature phase at $\beta=0.2391$, 
$0.23142$, $0.2275$, $0.2260$,$0.2240$ and $0.22311$ using a demon-algorithm 
implemented in the multispin coding technique. A detailed discussion of 
this algorithm is given in ref. \cite{algorithm}. 
The update of a single spin takes $46 \times 10^{-9}$ sec.  on a HP 735 and 
$21 \times 10^{-9}$ sec. on a DEC Alpha 250 workstation for $L=120$. 
For $\beta=0.22311$ the integrated 
autocorrelation time of the magnetization was $\tau_{int} = 81.(2.)$ in 
units of sweeps. 
 
We used cubical lattices with periodic boundary conditions and 
 a linear extension of about $20 \xi$. It should be noted that  
test runs  revealed that in contrast to the high  temperature phase 
a linear lattice size of $6 \xi$ is clearly not sufficient to obtain 
results close to the thermodynamic limit. Some informations on the simulations
are collected in tab.2.

We computed $\langle S_0 \; S_\tau  \rangle$ for all values of $\tau$ available 
on the finite lattice.  We evaluated $S_0 \; S_\tau$ for 
all three lattice directions
and all possible translations. The connected part was then obtained by 
subtracting the expectation value of the 
square of the magnetization 
$\langle \left( 1/V \; \sum_n s_n \right)^2 \rangle$. 

In order to obtain an estimate for the true correlation length we started
from the ansatz
\begin{equation}
\label{sing}
G(\tau) \;\; \propto \;\; \exp(-\tau/\xi) \;+\; \exp(-(L-\tau)/\xi) \;\;\;,
\end{equation}
where the last term takes into account the periodicity of the lattice. 
An effective correlation length $\xi_{eff}(\tau)$ is then computed by solving
the equation 
above for $\tau$ and $\tau+1$.
Ignoring the term $\exp(-(L-\tau)/\xi)$ , $\xi_{eff}(\tau)$
takes the form 
\begin{equation}
 \xi_{eff}(\tau)=\frac{1}{\ln(G(\tau+1))-\ln(G(\tau))};
\label{xieff}
\end{equation}
For $\tau > 3 \xi$ $\xi_{eff}(\tau)$ seems to stabilize within error bars. 
In tab.3 the results for $\tau = 3\xi$ are given. 
However it is important to stress that there might still
be systematic errors due to higher
excitations that are of the same  magnitude as the statistical error of 
$\xi_{eff}$.
Therefore a multi-exponential ansatz might be useful. 
We shall further
 discuss this point in sect. 4.4 below.

\vskip 0.4cm
We computed the second moment of the correlation function by 
\begin{equation}
\label{mu}
  \mu_2 = \sum_{\tau=1}^{\tau_{max}} \tau^2 G'(\tau)
+\sum_{\tau=\tau_{max}+1}^{\infty}\;\;\tau^2 \;\;\; G'(\tau_{max})
\;\; \exp(-(\tau-\tau_{max})/\xi_{eff}(\tau_{max}))
\end{equation}
where  $G'(\tau) = C_{eff}(\tau) \;\; \exp(-\tau/\xi_{eff}(\tau))$. 
Where again 
$C_{eff}(\tau)$ and $\xi_{eff}(\tau)$ are obtained from 
\begin{equation}
G(\tau) \;\; \propto \;\; \exp(-\tau/\xi) \;+\; \exp(-(L-\tau)/\xi) \;\;\;,
\end{equation}
inserting $\tau$ and $\tau+1$. 
As for the exponential correlation length we used $\tau_{max} = 3 \xi$ 
for the data
reported in tab.3 . Note that the systematic error introduced 
by the finite $\tau_{max}$ only affects the second term of 
eq. \ref{mu}, which is small
compared to the first one. Hence these systematic error can safely be ignored 
for our choice of $\tau_{max}$. 
The susceptibility is computed analogously.  The results are summarized 
in tab. 3.

\begin{table}[h]
\label{statlt}
\caption{\sl Statistics of the runs in the low temperature phase.}
\vskip 0.2cm

\begin{tabular}{|c|c|c|c|c|}
\hline
 $\beta$ & $L$ & measures& sweeps/measure& bits\\
\hline
0.2391 & 30&40000&25 &64\\
0.23142& 40&50000&25 &64\\
0.2275 & 50&50000&25 &64\\
0.2260 & 80&50000&25 &64\\
0.2240 &100&92000&25 &32\\
0.22311&120&124000&25&32\\
\hline
\end{tabular}
\end{table}

\begin{table}[h]
\label{spinlt}
\caption{\sl Results in the low temperature
phase}
\vskip 0.2cm
\begin{tabular}{|c|c|c|c|c|c|c|}
 \hline
 $\beta$ & $L$ &  $m$  & $E$  &  $\xi_{exp}$ & $\xi_{2nd}$   &  $\chi$ \\ 
 \hline
0.2391 & 30&0.667162(20) &0.553732(17)&1.2851(28)& 1.2335(15)& 4.178(3) \\ 
0.23142& 40&0.570306(16) &0.478046(12)&1.8637(45)& 1.8045(21)& 9.394(4) \\
0.2275 & 50&0.491676(14) &0.430364(10)&2.578(7)  & 2.5114(31)&18.706(10) \\
0.2260 & 80&0.449984(16) &0.409609(4) &3.103(7)  & 3.0340(32)&27.596(11) \\
0.2240 &100&0.372490(10) &0.378612(3) &4.606(13) & 4.509(6)  &61.348(34)\\
0.22311&120&0.320830(10) &0.362946(2) &6.208(18) & 6.093(9)  &112.60(7)\\
\hline
\end{tabular}
\end{table}

\subsection{Simulations in the high temperature phase.}

We simulated the Ising spin model
in the 
high temperature phase using the single cluster algorithm \cite{ulli}. 
The time slice correlation function was determined using the cluster improved 
estimator. 
Finite size effects are less important in this phase and some preliminary test 
showed that lattice sizes greater than $6\xi$ are enough to keep them under
control. This is clearly visible in the data of tab.4 where we report a test
at $\beta=0.21931$.
\begin{table}[h]
\label{spinht0}
\caption{\sl Test for finite size corrections at $\beta=0.21931$.}
\vskip 0.2cm
\begin{tabular}{|c|c|c|c|c|c|c|c|}
\hline
$\beta$ &$L$&  stat   & $\xi_{exp}$ & $\xi_{2nd}$ & $E$  & $\chi$    \\
\hline
0.21931&40 &50000x1000 &8.701(4)&8.691(5)&0.313986(24) &303.23(30)  \\
0.21931&50 &50000x1000 &8.747(4)&8.741(5)&0.313870(18) &307.14(31)  \\
0.21931&60 &50000x1000 &8.754(6)&8.750(5)&0.313811(15) &307.79(31)  \\
0.21931&70 &50000x1000 &8.758(4)&8.751(5)&0.313823(14) &307.95(31)  \\
0.21931&80 &50000x1000 &8.766(5)&8.760(5)&0.313849(14) &308.58(31)  \\
\hline
\end{tabular}
\end{table}

Also the determination of the correlation length in the high temperature phase
turns out to be much easier than in the low temperature phase.  The effective
correlation length approaches a  plateau quite quickly and the true correlation
length is well approximated by $\xi_{eff}$ at a self-consistently chosen 
distance $\tau \approx \xi_{eff}$. 
This behaviour of  $\xi_{eff}$ also implies that the 
difference  between
the second moment correlation length
and the true correlation length is much smaller that in the low temperature
phase. 
We have chosen the $\beta$ values in the high temperature phase
such that $\beta_c-\beta=\beta_{low}-\beta_c$, where $\beta_{low}$ are the 
inverse temperature used in the simulations of low temperature phase. 
The reason for this choice will be clear in the following section. 
 The uncertainty
of $\beta_c$ virtually does not affect the following analysis.   
The simulation results are summarized in tab.5.
\begin{table}[h]
\label{spinht}
\caption{\sl Results in the high temperature
phase}
\vskip 0.2cm
\begin{tabular}{|c|c|c|c|c|c|c|c|}
\hline
$\beta$ &$L$&  stat   & $\xi_{exp}$ & $\xi_{2nd}$ & $E$  & $\chi$    \\
\hline
0.20421&20 &50000x500 & 2.363(1)&2.346(1)&0.262928(49) & 25.255(16) \\
0.21189&30 &50000x500 & 3.477(2)&3.465(2)&0.284663(35) & 52.09(4)  \\
0.21581&40 &50000x500 & 4.864(3)&4.854(3)&0.298366(28) & 98.90(10)  \\
0.21731&50 &50000x800 & 5.892(3)&5.885(3)&0.304493(20) &143.04(12)  \\
0.21931&80 &50000x1000 &8.766(5)&8.760(5)&0.313849(14) &308.58(31)  \\
0.22020&100 &50000x1200&11.884(9)&11.877(7)&0.318742(11)& 557.57(61) \\
\hline
\end{tabular}
\end{table}
Recently  in ref. \cite{kim0} Monte Carlo results for the second moment correlation
length and the magnetic susceptibility were reported. Interpolation of their results,
using the scaling ansatz, to our $\beta$-values leads to results consistent with 
ours. One has to note however, that our statistical errors are at least 3 times 
smaller than those of ref. \cite{kim0} in the common $\beta$-range. 

\section{Analysis of the results}
\subsection{Magnetization}
A very precise Montecarlo study of the behaviour of the magnetization
in the Ising model can be found in~\cite{tb}. In particular, in~\cite{tb}
it was shown that the magnetization values obtained from the montecarlo
simulations were well described by the following empirical approximation:
\eq
m(\beta)=t^{0.32694109}(1.6919045-0.34357731~t^{0.50842026}-
0.42572366~t)
\label{eqtp}
\en
with $t$ given by eq.(\ref{tbeta}).
Since our values of~$\beta$ are inside the region of validity of this
approximation, it is  interesting to compare also our magnetization values 
with eq.(\ref{eqtp}). Notice as a side remark, that our estimates for the
magnetization are in general more precise of those reported in~\cite{tb} (where
however a much larger number of $\beta$ values was studied).
 This comparison is reported in tab.6, together with
the estimates obtained with a double biased IDA
analysis of the series published in~\cite{ge2}.

\begin{table}[h]
\label{spinht2}
\caption{\sl Comparison of our Monte Carlo results for the  magnetization
with eq.~\ref{eqtp}  and with double biased IDA's.}
\vskip 0.2cm
\begin{tabular}{|c|c|c|c|}
\hline
  $\beta$ &  our MC      &  eq. 10 of \cite{tb}& biased IDA \\
\hline
   0.2391 &  0.66716(2)  &  0.667143 & 0.667151(3)(1) \\    
   0.23142&  0.570306(16)&  0.570279 & 0.570300(16)(1) \\
   0.2275 &  0.491676(14)&  0.491645 & 0.49167(4)(1) \\ 
   0.2260 &  0.449984(16)&  0.449953 & 0.44999(7)(1) \\
   0.2240 &  0.372490(10)&  0.372471 & 0.37253(15)(2) \\
   0.22311&  0.320830(10)&  0.320809 & 0.3209(2)(1) \\
\hline
\end{tabular}
\end{table}

It is interesting to notice that our values are always in perfect agreement
with those obtained from the series expansions, and that our results become
more precise than the strong coupling ones starting from  $\xi\sim 2$.

The agreement with the data of ref~\cite{tb} is also very good. Even if our 
values are systematically slightly higher than those of eq.(\ref{eqtp}), they 
are well inside the error bars reported in tab.1 of ref~\cite{tb}.

\subsection{Susceptibility}
In tab.7 we report the comparison of our data on the susceptibility in the low
temperature phase with a double biased IDA analysis of the series published
in~\cite{ge2}. The agreement is again very good and as in the previous case 
our results become
more precise than the strong coupling ones starting from  $\xi\sim 2$.
\begin{table}[h]
\label{susceptibility}
\caption{\sl Comparison of our Monte Carlo results for the susceptibility
 with double biased IDA's.}
\vskip 0.2cm
\begin{tabular}{|c|c|c|}
\hline
  $\beta$ &  our MC      &   biased IDA \\
\hline
   0.2391 &  4.178(3)  &   4.1801(16) \\    
   0.23142&  9.394(4)  &   9.401(20) \\
   0.2275 &  18.706(10)&   18.76(15) \\ 
   0.2260 &  27.596(11)&   27.67(40) \\
   0.2240 &  61.348(34)&   61.8(2.7) \\
   0.22311&  112.60(7) &   114.6(10.5) \\
\hline
\end{tabular}
\end{table}

\subsection{Second moment correlation length}
In tab.8 we report the comparison of our data on the susceptibility in the low
temperature phase with a double biased IDA analysis of the series published
in~\cite{at1}. Also in this case the agreement is very good. 
In addition we give Montecarlo results of ref. \cite{kim} for which the 
$\beta$-value
matches with ours. One has to note that the results of ref. 
\cite{kim} were obtained
with $L \approx 4.7 \xi $ and $L \approx 7.8 \xi $ for $\beta=0.226$ and 
$\beta=0.224$ respectively. 

\begin{table}[h]
\label{corr2}
\caption{\sl Comparison of our Monte Carlo results for the second moment
correlation length with double biased IDA's.}
\vskip 0.2cm
\begin{tabular}{|c|c|c|c|}
\hline
  $\beta$ &  our MC      &MC of ref \cite{kim}&   biased IDA \\
\hline
   0.2391 &  1.2335(15)  &            & 1.2358(16) \\    
   0.23142&  1.8045(21)  &            & 1.803(5)  \\
   0.2275 &  2.5114(31)  &            & 2.509(11) \\ 
   0.2260 &  3.0340(32)  &  3.22(1)   & 3.034(16) \\
   0.2240 &  4.509(6)    &  4.61(6)   & 4.493(30) \\
   0.22311&  6.093(9)    &            & 6.084(46) \\
\hline
\end{tabular}
\end{table}

\subsection{Exponential correlation length}
As we mentioned above, in the low temperature phase the evaluation of the
exponential correlation length is much more delicate than in the
high temperature phase. In particular, we know from the fact that the ratio
 $\frac{\xi}{\xi_{2nd}}$ is significantly different from 1 and
from~\cite{us} that in this region the spectrum is very rich and that nearby
states exist that could contaminate the measure of $\xi$. This is exactly the
situation discussed in sect.3.1 above and accordingly  we may 
expect some systematic error in $\xi_{eff}$. Since the presence of nearby
masses is a rather common 
situation 
in the broken symmetry phases of statistical mechanical models and since,
notwithstanding this, the estimator $\xi_{eff}$ is commonly used also in this
cases, we have decided to devote this section to a detailed analysis of this
problem. 
We can explicitly see that
 $\xi_{eff}$ evaluated according to eq.(\ref{xieff}) is not a good
estimator of the true correlation length by comparing our estimates with those
of ref.~\cite{us} (see tab.9).
 In ref.~\cite{us} we computed the 
glueball spectrum of the $Z_2$ gauge theory in 3 dimensions. In $d=3$ 
 the spin and gauge Ising models are related by duality and the inverse of the
 $0^+$ glueball mass exactly coincides with exponential correlation length of
 the spin Ising model. In ref. \cite{us}
 we used a variational approach, using 27 different wilson loops as
operators, to obtain a faster convergence of $\xi_{eff}$. In this way each mass
of the spectrum was driven in a different channel, and practically 
no contamination from
higher states was present.
The results of~\cite{us} are comparable in statistical accuracy with the
ones presented here. It is easy to see looking at tab.9 that the
values of $\xi_{eff}$ 
obtained here are systematically smaller, in average by a factor of  
$0.995(2)$. This shows, as expected, that the single exponential ansatz is 
problematic in this case and that a multi-mass ansatz is needed. 
In order to have an independent test of this fact
we tried to fit our data at $\beta=0.224 $ for the correlation function 
 with the following 3-mass ansatz
\begin{equation}
G(\tau) \;\; \sim \;\; c_1 \; \exp(-\tau/\xi_1) + c_2 \; \exp(-\tau/\xi_2)
	     +  c_3 \; \exp(-\tau/\xi_3)~~~ .
\end{equation}
However, the main problem of
such multi-mass fits is that they are in general rather unstable
under variation of 
the fit-range. This was also the case with our fits.
Therefore we fixed the values of $\xi_2=2.50$ and $\xi_3=1.70$ 
found in ref. \cite{us}. We found the following values:
$c_2/c_1 \approx 0.1 $ and $c_3/c_1 \approx 0.04$,  when 
$\tau$'s in the range of $5$ to $25$ are included in the fit. 
Notice however that even in this case the results were still rather unstable,
and for this reason we cannot give reliable error bars for our estimates. 
 Assuming that 
$G(\tau)$ is well described by the three mass-ansatz and using  our
estimates for 
$c_2/c_1$ and $c_3/c_1$ we obtain $\xi_{eff}(3 \xi)=0.993 \xi$ which is indeed
consistent with our result above.

It is also interesting to insert our estimate for $c_2/c_1$ and $c_3/c_1$
into eq. \ref{e24}. 
We obtain $\xi/\xi_{2nd} =1.024$. 

The situation is much simpler in the high temperature phase, where
$\xi_{2nd}\sim\xi$, no nearby masses are present and $\xi_{eff}$ is a good
estimator of the true correlation length.

\begin{table}[h]
\label{comxi}
\caption{Comparison of the results for the exponential 
correlation length with those
obtained for the $Z_2$ gauge model ($\tilde\beta$ denotes the dual of $\beta$)}
\vskip 0.2cm
\begin{tabular}{|l|c|l|l|}
\hline
$\tilde\beta$ & $\beta$& $\xi_{gauge}$ & $\xi_{eff}$ \\
\hline
0.72484  & 0.23910  &  1.296(3) & 1.2851(28)\\
0.74057  & 0.23142  &  1.864(5) & 1.8637(45)\\
0.74883  & 0.22750  &  2.592(5) & 2.578(7) \\
0.75202  & 0.22600  &  3.135(9) & 3.103(7) \\
0.75632  & 0.22400  &  4.64(3)  & 4.606(13)\\
\hline
\end{tabular}
\end{table}

\section{Universal amplitude ratios}
The standard approach to evaluate the amplitude ratios
 is to fit the data
obtained for both phases separately with the expected scaling law, and then 
take the ratio of the amplitudes obtained from the fits. 

However the bias introduced by the  uncertainty in the critical exponent
 can be avoided by directly studying the ratio as a function of the reduced
 temperature. This is the reason for which we carefully chose the couplings so
 as to have the same differences $\Delta \beta\equiv|\beta-\beta_c|$ 
in the two phases.
 To better explain our approach let us study as an example the amplitude
 ratio $\Gamma_\chi$. From the data reported in tab.3 and 5
we can compute the ratios of susceptibilities in the low and high temperature 
phase as a function of $\Delta\beta$.
\begin{equation}
\Gamma_{\chi}(\Delta \beta) = \frac{\chi(\beta_c-\Delta \beta)}
			      {\chi(\beta_c+\Delta \beta)}
\end{equation}
As $\Delta \beta$ goes to $0$ we expect $\Gamma_{\chi}(\Delta \beta)$ to 
converge to the amplitude ratio $C_+/C_-$. However the approach to this 
critical value is rather non-trivial. A naive implementation of the  
scaling hypothesis would suggest that the data thus obtained should be constant
within the error, but it is easy to see, by looking at the data in tab.10 that
this is not the case. There are in fact two source of corrections.
 The fact that observables in both phases
are involved in the ratio tells us that we must expect a correction
proportional to $\Delta\beta$. Moreover we certainly expect a ``correction
to scaling'' contribution proportional to $\Delta \beta^{\theta}$.
In the example of $\Gamma_\chi$ the need of such corrections is clearly 
evident. Looking at the data in tab.10 we see that
the violations of scaling are much larger than our statistical errors. 
Even for the smallest values of $\Delta \beta$ we see no stabilization 
of $\Gamma_{\chi}$ within error bars. 
Following the above discussion we  fitted the data of tab.10 with the law
\begin{equation}
\label{fit2}
\Gamma_{\chi}(\Delta \beta) \; = \; C_+/C_- \;+\; a_0 \;\Delta \beta^{\theta}
				 \; +\; a_1 \; \Delta \beta \;\;\; .
\end{equation}
where we  assumed that there is no other correction to scaling
exponent $\theta'$ between $\theta$ and $1$. 
The results of these fits for the various ratios in which we are interested are
reported in tab.11. The fact that we always find rather low $\chi^2_{red}$
 strongly supports the correctness of 
 the above assumption.

\begin{table}[h]
\label{datafit}
\caption{The various ratios as  functions of $\Delta\beta$}
\vskip 0.2cm
\begin{tabular}{|l|l|l|l|l|l|}
\hline
$\Delta \beta$  &  $\Gamma_{\chi}$ &  $\Gamma_{\xi}$ & $u$ 
&$\Gamma_c$ & $\frac{\xi}{\xi_{2nd}}$ \\
\hline
0.01745  &       6.044(5) &  1.902(2) &  15.00(6) &  4.394(9) &  1.042(3)\\
0.00977  &       5.546(5) &  1.920(3) &  14.75(5) &  3.850(10)&  1.033(3)\\
0.00585  &       5.283(3) &  1.932(3) &  14.66(6) &  3.577(10)&  1.026(4)\\
0.00435  &       5.182(5)(1)&  1.939(3)&  14.64(5)&  3.466(8) &  1.0227(34)\\
0.00235  &       5.027(6)(2) &  1.942(3) &  14.47(6)&  3.308(9)&  1.0215(42)\\
0.00146  &       4.947(6)(3) &  1.948(3)(1) &  14.51(7)&  3.233(9)(1)&  
1.0188(45)\\
\hline
\end{tabular}
\end{table}

Few comments are in order at this point.
\begin{description}
\item{a1]} 
When dealing with combinations of observables all in
the same phase we don't need a correction to scaling term proportional to
$\Delta\beta$. This is the case of the coupling constant $u$ and of the
ratio $\xi/\xi_{2nd}$. In this case we fitted with the 
law\footnote{Notice however that, hidden in the  $\Delta \beta$ correction there
should also be a term proportional to $\Delta \beta^{2\theta}$ which, due to
the fact that $\theta\sim 1/2$ is essentially indistinguishable from 
$\Delta \beta$. The correction $\Delta \beta^{2\theta}$ should be present 
also in the case in which all the observables belong to the same phase, thus
suggesting to use also in this case the fit eq.(\ref{fit2}). It turns out
however that such a $\Delta \beta^{2\theta}$ correction, if present, has a
negligible amplitude and for this reason we confined ourselves to the fit
(\ref{fit}).}:
\begin{equation}
\label{fit}
u(\Delta \beta) \; = \; u^* \;+\; a_0 \;\Delta \beta^{\theta}
\end{equation}

\item{a2]} The error due to the specific choice of $\beta_c$ is always very
small. We have reported its value in the data of tab.10 only 
when it is not negligible. 
In these cases the number in the first bracket gives the 
 statistical errors of our data, while the second takes into
account the uncertainty of the inverse critical temperature. 

\item{a3]}
In the last row of tab.11 we have reported our final results for the various
ratios. The corresponding errors take also into account the uncertainty in the
index $\theta$ and are thus slightly larger than those extracted by the fits.

\item{a4]}
In the results reported in tab.11 we always fitted only the last five 
values of $\Delta \beta$, and systematically discarded the data at
$\Delta \beta = 0.01745$\footnote{This is not due to the fact that adding 
this data we had a poorer fit, on the contrary we checked that
 that for all the ratios the fit keeping {\sl all the six } data was always
 equally good. The reason of our choice is that we tried to confine ourselves to
 the narrowest possible
 region near the critical point compatible with a reasonable precision for the
 results. This allows us to trust in our assumption of neglecting other 
 possible, unknown, corrections to scaling which are certainly present but
 hopefully negligible in this range.
It is a remarkable consequence of the high precision of our montecarlo estimates
that we can still extract meaningful results by using only five values of
$\Delta\beta$.}.

\item{a5]} A particular care must be devoted to the study of the ratio
 $\frac{\xi}{\xi_{2nd}}$ . 
It is possible to prove that if
higher masses exist in the theory (and this is the case in both
phases of the Ising model) then $\frac{\xi}{\xi_{2nd}}$
must certainly be larger than 1. This is a consequence of 
 eq.(\ref{e24}) and of the fact that the coefficients $c_i$ which appear in it
 must be positive (see eq. (9) and (10) of ref.~\cite{us} for a proof of this
 last statement). 

In the high temperature phase $\frac{\xi}{\xi_{2nd}}$ is almost compatible with
1 (within the errors), and we can only use our data to set an upper bound for
its value which, looking at the data with the largest correlation length, 
can be safely chosen to be $\frac{f_+}{f_{+,2nd}}<1.0006$.

On the contrary in the $\beta>\beta_c$  phase the quantity 
$\frac{\xi}{\xi_{2nd}}-1$
is much larger than the error bars, and can be measured rather precisely. 
 The data in the
last column of tab.10 and 11 refer to this case and use $\xi_{eff}$ (defined in
sect.4.4) as estimator of $\xi$. Hence we must add to
 the result of the fit (which is reported in the last row of tab.11:
$\frac{f_-}{f_{-,2nd}}=1.009(5)$ )
 the
contribution due to the systematic underestimation $\Delta\xi\sim 0.007$
 discussed in sect.4.4 above. Taking into account also this correction
 we quote as our final result $\frac{f_-}{f_{-,2nd}}=1.017(7)$.
It is interesting to notice that this result agrees within the errors with 
the value $\frac{f_-}{f_{-,2nd}}\sim 1.024$ obtained in sect.
4.4 by inserting in eq.(\ref{e24}) our estimates for $c_2/c_1$ and $c_3/c_1$
and the values of the two nearby (inverse) masses $\xi_2$ and $\xi_3$ 
extracted from ref.~\cite{us}. Finally we can directly estimate the ratio by
using the unbiased data for $\xi$ obtained in ref.~\cite{us} and reported in
tab. 9. The only problem is that these data are slightly less precise and that
the lowest value of $\beta$ is missing. The resulting estimate for the ratio:
$\frac{f_-}{f_{-,2nd}}= 1.029(11)$ is thus affected by a larger error. Also in
this case we find a good agreement within the errors with our final result
$\frac{f_-}{f_{-,2nd}}=1.017(7)$

\end{description}

\begin{table}[h]
\label{results}
\caption{Results of the fits according to eq.(\ref{fit2}) and (\ref{fit})
(see also the comment (a5) above).}
\vskip 0.2cm
\begin{tabular}{|l|l|l|l|l|l|}
\hline
  &  $\Gamma_{\chi}$ &  $\Gamma_{\xi}$ & $u$ 
&$\Gamma_c$ & $\frac{\xi}{\xi_{2nd}}$ \\
\hline
$\chi^2_{red}$  &  0.52 &  0.43 &  0.52 &  0.01 &  0.20\\
C.L.  &            59\%(2) &  65\% &  67\%&  98\% &  90\%\\
$a_0$  &           3.5(8) &  -0.04(50) &  4.8(1.2) &  3.0(1.5)&  0.24(8)\\
$a_1$ &            46.9(5.4) &  -2.9(3.4) &   &  54(10)&  \\
\hline
\vbox{\hbox{final}
\vskip 0.1cm
\hbox{result}}  &  4.75(3)&  1.95(2)&  14.3(1)&  3.05(5) &  1.009(5)\\
\hline
\end{tabular}
\end{table}

\subsection{Comparison with other existing estimates}
In tab.14 we have compared our
 results with those obtained with other methods.
 Let us briefly comment on this comparison:
\begin{description}
\item{b1]} There are three possible approaches to the
evaluation of the amplitude ratios. Montecarlo simulations (``MC'' in tab.14),
low and high temperature series expansions (``HT,LT'', in tab.14) and field
theoretic methods. In this last case two different approaches are possible. 
The first one consists in looking at the $\epsilon$ expansion of the $\phi^4$
theory around four dimensions (``$\epsilon$-exp.'' in tab.14). The second one
consists in looking directly to the $\phi^4$ theory in three dimensions
(``d=3'' in tab.14). For
a detailed discussion of these approaches see for instance ref.~\cite{zjbook}.
Let us also mention for completeness that an independent, interesting method to
evaluate the ratio of specific heat amplitudes (which we do not study in this
paper) by looking at the distribution of the zeroes of the partition function
was proposed and applied to the Ising model in~\cite{ipz,m84}. 

\item{b2]} The results of ref.~\cite{blz,an} for $\Gamma_\chi$ and $\Gamma_\xi$ were
obtained with a careful resummation of  two loop
$\epsilon$-expansions. On the contrary, the $\epsilon$-expansion
for the exponential $\xi$ (which is needed to obtain the value of 
$f_-/f_{-,2nd}$ which is reported in the last column of tab.14) is known only
at one loop, hence the value $f_-/f_{-,2nd}\sim 1.005$ of~\cite{blz} must only be
considered as indicative. Later the $\epsilon$-expansion for $\Gamma_\chi$ was
extended up to $\epsilon^3$ and the value of~\cite{an} corresponds to the 
Pad\`e
resummation of such series. Recently in~\cite{gz} this result has been further
improved by using the  parametric representation of the 
 equation of state of the theory.

\item{b3]} The $d=3$ approach originates from a suggestion of
Parisi~\cite{Parisi}. While the results of~\cite{gz,bbmn} make use only of
 series expansions obtained in the symmetric phase of the theory,
 in~\cite{munster1,munster2} a three loop calculation, directly performed in
 the low temperature phase, was used. 
 In this last case a crucial role is played by the
 low temperature renormalized coupling constant $u$ evaluated at the critical
 point. We shall further comment on this point later.

\item{b4]}
 The estimates for the amplitude ratios obtained with the use of low and high 
 temperature series expansion reported in tab.14 are mainly taken from 
 ref~\cite{lf}. They were obtained with the use of IDA's on the series 
 reported in tab.12, where we used the standard notations: $v\equiv th(\beta)$ 
and  $u\equiv e^{-4\beta}$ (not to be confused with the low temperature
coupling constant in the $d=3$ $\phi^4$ theory!)
for high and low temperature series respectively.
 Recently these data have been reanalyzed in~\cite{zf} (using the same series)
 leading to essentially the same results.  Notice however that in these last 
 years new longer series have been constructed in the low temperature phase. 
 These are the series that we have used in the previous sections to test our
 montecarlo results. It would be very interesting to see if these new series
 can lead to improved estimates of the amplitude ratios. In particular it would
 possible now to analyze also  the ratio $\xi/\xi_{2nd}$ which
 was previously unaccessible, since the two series  for $\xi$ and  for
 $\xi_{2nd}$ start to be different only at the order $u^6$ 
\begin{table}[h]
\label{series1}
\caption{\sl Some informations on the series used in ref.[9]}
\vskip 0.2cm
\begin{tabular}{|c|c|c|c|}
\hline
ref.& year & observable & length \\
\hline
\cite{gs}& 1979 & HT/~$\chi$ & $v^{18}$ \\
\cite{mjw}& 1969 & HT/~$\xi_{2nd}$ & $v^{12}$ \\
\cite{gsee}& 1979 & LT/~$\chi$ & $u^{20}$ \\
\cite{tf}& 1975 & LT/~$\xi_{2nd}$ & $u^{15}$ \\
\cite{tf}& 1975 & LT/~$\xi$ & $u^{7}$ \\
\cite{gsee}& 1979 & LT/~$m$ & $u^{21}$ \\
\hline
\end{tabular}
\end{table}

\begin{table}[h]
\label{series2}
\caption{\sl Some informations on the low temperature series used in this
paper.}
\vskip 0.2cm
\begin{tabular}{|c|c|c|c|}
\hline
ref.& year & observable & length \\
\hline
\cite{ge2}& 1993 & LT/~$\chi$ & $u^{32}$ \\
\cite{at1}& 1995 & LT/~$\xi_{2nd}$ & $u^{23}$ \\
\cite{at2}& 1995 & LT/~$\xi$ & $u^{15}$ \\
\cite{ge2}& 1993 & LT/~$m$ & $u^{32}$ \\
\hline
\end{tabular}
\end{table}

\item{b5]}
Some of the data reported in tab.14 (those which are underlined) have been
obtained combining separate amplitudes reported by the authors, thus their
errors are most probably overestimated.

\item{b6]}
It has been recently reported in~\cite{cprv}
 an estimate for the ratio $\frac{f_+}{f_{+,2nd}}$ obtained 
  with a strong coupling expansion to 15th order of the correlation function
$G(\tau)$. The result is  
$\frac{f_+}{f_{+,2nd}}=1.00023(5)$ which  agrees with our bounds
$1.<\frac{f_+}{f_{+,2nd}}<1.0006$.

\end{description}

\begin{table}[h]
\label{literature2}
\caption{\sl Results for the amplitude ratios reported 
 in the literature}
\vskip 0.2cm
\begin{tabular}{|c|c|c|c|c|c|c|c|}
\hline
ref.& year & method & $\frac{C_+}{C_-}$ & $\frac{f_{+,2nd}}{f_{-,2nd}}$
 & $u^*$ & $\frac{C_+}{f^3_{+,2nd}B^2}$ 
& $\frac{f_-}{f_{-,2nd}}$ \\
\hline
\cite{blz,ah}& 1974 &$\epsilon$-exp & $\sim 4.8$ & $\sim 1.91$
& & &$\sim 1.005$ \\  
\cite{an}& 1985 &$\epsilon$-exp & $\sim 4.9$ & & & & \\  
\cite{gz}& 1996 & $\epsilon$-exp &  4.70(10) & & & &\\  
\cite{bbmn}& 1987 &$d=3$ & 4.77(30) & & &\underline{3.02(8)} & \\
\cite{gz}& 1996 &$d=3$ & 4.82(10) & & & & \\
\cite{munster2}& 1996 &$d=3$ & 4.72(17) & 2.013(28) & 
\underline{15.1(1.3)} & & \\
\cite{lf}& 1989 & HT,LT & 4.95(15) & 1.96(1) 
&\underline{14.8(1.0)} &\underline{3.09(8)} & \\
\cite{siep}& 1993 & HT,LT & & & 14.73(14)& & \\
\cite{kielxi}& 1994 & MC &\underline{5.18(33)}& 2.06(1) &
\underline{17.1(1.9)} & \underline{3.36(23)} & \\
This work &  1996 & MC & 4.75(3) & 1.95(2) & 14.3(1) & 3.05(5) & 1.017(7) \\
\hline
\end{tabular}
\end{table}

\subsection{Comparison with an effective potential model.}
It has been recently proposed to study the critical properties of the three
dimensional Ising model by constructing the effective potential of the
corresponding quantum field theory. This effective potential is constructed
simulating  the model for various values of the external magnetic field.
This program was carried on in~\cite{tsypin1} for the high temperature phase of
the model and was recently extended to the broken symmetric phase
in~\cite{tsypin}.
The main result is that in the effective
potential, besides the expected $\phi^4$ term  also a $\phi^6$ term is present.
It is interesting to test this model with our high precision results.
Fortunately one of the values of $\beta$ studied in~\cite{tsypin}: 
$\beta=0.2260$ exactly coincide with one of our values thus allowing a detailed
comparison. This comparison is reported in tab.15, where in the third column we
have reported the values of the various observables directly measured at
$\beta=0.2260$ (tab.1 of~\cite{tsypin}) while in the last column we have
reported the same observables obtained from what was considered
 in~\cite{tsypin} as
the most successful fitting procedure (tab.2 of~\cite{tsypin}).
Notice that $\chi$ is the inverse of $V''$ and that $u=3G$ 
in the notations of~\cite{tsypin}. Our results are in general one order of
magnitude more precise than those of ref.~\cite{tsypin}.
 It is interesting to see that both $m$ and
$\chi$ are in rather good agreement with our data. The only strong disagreement
is in the value of $\xi_{2nd}$ and, as a consequence of this, in $u$. 
Most likely this disagreement is only due to the too small lattices studied in 
ref.~\cite{tsypin} (the lattice size is reported in the last line of tab.15)
and does not imply that the approach proposed in~\cite{tsypin} is wrong.

\begin{table}[h]
\label{tsypin2}
\caption{\sl Comparison with Tsypin's results.}
\vskip 0.2cm
\begin{tabular}{|c|c|c|c|}
\hline
Observable & This work & Tab.1 of~\cite{tsypin} &  Tab.2 of~\cite{tsypin}  \\
\hline
$m$ & 0.449984(16)  & 0.44975(17) & 0.44975(17) \\
$\chi$ & 27.596(11) & 27.397(75) & 27.586(198)\\
$\xi_{2nd}$ & 3.0340(32) & 2.946(6) & 2.956(10) \\
$u$ & 14.64(5) & 15.90(9) & 15.84(28)  \\
$L$& 80 & 30 & 30\\
\hline
\end{tabular}
\end{table}

\subsection{Comparison with experimental data}

 The experimental data reported in tab.16 refer to the three most
important experimental realization of the Ising universality class, namely 
binary mixtures (bm),  liquid-vapour transitions (lvt) and uniaxial
 antiferromagnetic systems (af). 
It is important to notice that these realizations are not on the
same ground. Antiferromagnetic systems are particularly apt to measure the
$C_+/C_-$ and $f_{+,2nd}/f_{-,2nd}$ ratios, while for
 the liquid vapor transitions 
the $\Gamma_c\equiv R_c/R_\xi^3$ combination is more easily accessible.
Finally, in the case of binary mixture all the three ratios can be rather
easily evaluated. 
Even if obtained with very different experimental setups all these estimates 
qualitatively agree among them and this is certainly one of the most remarkable
experimental evidences of universality. When looking in more detail at
the various results one can see a residual small spread among them (even if
 in general the various estimates are compatible
 within the quoted experimental uncertainties). 
This spread is mainly due
to the presence of correction to scaling terms whose amplitudes
vary as the experimental realizations are changed
and that are difficult to control. 
Thus some care is needed to compare
these experimental data with theoretical estimates.
The common attitude is to assume
 that the above systematic
errors are randomly distributed and to take
 the weighted mean of the various experimental results. 
\begin{table}[h]
\label{literature3}
\caption{\sl Experimental estimates for some amplitude ratios}
\vskip 0.2cm
\begin{tabular}{|c|c|c|c|}
\hline
exp. setup  & $\frac{C_+}{C_-}$ & $\frac{f_{+,2nd}}{f_{-,2nd}}$
  & $\frac{C_+}{f^3_{+,2nd}B^2}$  \\
\hline
 (bm)  & 4.4(4) & 1.93(7) & 3.01(50) \\
 (lvt) & 4.9(2) & &  2.83(31)\\
 (af)  & 5.1(6) & 1.92(15)  &   \\
 (all of them) &4.86(46) & 1.93(12) & 2.93(41)  \\
This work &   4.75(3) & 1.95(2)  & 3.05(5)  \\
\hline
\end{tabular}
\end{table}

Following this line  
we have reported in tab.16  the weighted means (together with,
 in parenthesis, the standard deviations),
 of the experimental results reported in~\cite{ahp}. 
In the first three rows we have studied separately the three different 
realizations of the universality class while in the fourth row all the
experimental data at disposal are analyzed together. In the last row we have
reported our results.

In the case of the $f_{+,2nd}/f_{-,2nd}$ ratio ( for which, as we have seen, 
 some of the present theoretical or montecarlo estimates disagree)
 we have listed, for a more detailed comparison, all the available 
experimental data in tab.17. In this table we denote with 
``N-H'' the nitrobenzene~--~n-hexane binary mixture, and with
 ``I-W'' the one obtained by mixing isobutyric acid and water.
 A much more detailed account of the various 
experimental estimates can be found in~\cite{ahp}.

\begin{table}[h]
\label{literature4}
\caption{\sl Experimental estimates for the $\frac{f_{+,2nd}}{f_{-,2nd}}$
ratio}
\vskip 0.2cm
\begin{tabular}{|c|c|c|c|}
\hline
Ref. & year & exp. setup &  $\frac{f_{+,2nd}}{f_{-,2nd}}$ \\
\hline
\cite{snhl} & 1971  & (af),${\rm MnF_2}$ & 1.7(3) \\
\cite{hsg} & 1972  & (af),${\rm FeF_2}$ & 2.06(20) \\
\cite{cc} & 1980  & (af),${\rm CoF_2}$ & 1.93(10) \\
\cite{zbb} & 1983  & (bm),N-H & 1.9(2) \\
\cite{htkk} & 1986  & (bm),I-W & 2.0(4) \\
This work &  1996 & MC & 1.95(2)   \\
\hline
\end{tabular}
\end{table}

\section{Conclusions}
We have estimated various universal amplitude ratios in the case of the three
dimensional Ising model. Our final results are:
\eq
 \frac{C_+}{C_-}=4.75(3)~~~~~~~~~~
 \frac{f_{+,2nd}}{f_{-,2nd}}=1.95(2)~~~~~~~~~~~
 \frac{f_-}{f_{-,2nd}}=1.017(7)
\nonumber
\en
\eq
 u^*\equiv\frac{3~C_{-}}{f^3_{-,2nd}B^2}=14.3(1)~~~~~~~~
 \frac{C_{+}}{f^3_{+,2nd}B^2}=3.05(5)~~~~~.
\nonumber
\en
Our results are in general in good agreement with other estimates of the same 
quantities obtained with field theoretical methods or with high/low
temperature series. 
The main discrepancy that we have found is with the Montecarlo results of 
ref.~\cite{kielxi}
and with some of the results of ref.~\cite{tsypin}. 
It must also be noticed that our result $ \frac{f_{+,2nd}}{f_{-,2nd}}=1.95(2)$
 only marginally agrees with that of~\cite{munster2}: 2.013(28). However, as we
 mentioned above, this result depends on the value of the coupling constant 
 $u^*$ which is an external input in the calculations of~\cite{munster2}. 
By plugging our value of  $u^*$ in the perturbative expansion
of~\cite{munster2} we find a lowering of the ratio
 with respect to~\cite{munster2}. This lower
result: $ \frac{f_{+,2nd}}{f_{-,2nd}}=1.99(2)$ agrees  not only with the 
strong/weak coupling result~\cite{lf}, but also with our estimate.
Finally it is important to notice that our results are also in reasonable
agreement with the experimental ones.

\vskip 1cm
{\bf  Acknowledgements}

We thank F.Gliozzi, K.Pinn, P.Provero and S.Vinti for many helpful 
discussions. 
This work was partially supported by the 
European Commission TMR programme ERBFMRX-CT96-0045.

\end{document}